\documentclass[prc]{revtex4}

\usepackage{graphicx}
\usepackage{amsmath}
\usepackage{amssymb}

\begin{document}

\title{\bf Power counting for three-body decays of a near-threshold state}
\author{Mohammad H. Alhakami}
 \altaffiliation{Permanent address: KACST, PO Box 6086,
Riyadh 11442, Saudi Arabia}
\author{Michael C. Birse}
\affiliation{Theoretical Physics Division, School of Physics and Astronomy, 
The University of Manchester, Manchester M13 9PL, UK}

\date{\today}

\begin{abstract}
We propose a new power counting for the effective field theory describing
a near-threshold state with unstable constituents, such as the $X(3872)$ meson.
In this counting, the momenta of the heavy particles, the pion mass and 
the excitation energy of the unstable constituent---the $D^*$ in the case 
of the $X$---are treated as small scales, of order $Q$. The difference $\delta$
between the excitation energy of the $D^*$ and the pion mass is smaller 
than either by a factor $\sim 20$. We therefore assign $\delta$ an order $Q^2$ 
in our counting. This provides a consistent framework for a double expansion in
both $\delta/m_\pi$ and the ratio of $m_\pi$ to the high-energy scales
in this system. It ensures that amplitudes have the correct behaviour at the
three-body threshold. It allows us to derive, within an effective theory, 
various results which have previously been obtained using physically-motivated
approximations.
\end{abstract}

\pacs{}

\maketitle

The $X(3872)$ meson has provided a puzzle since it was discovered by the Belle
collaboration \cite{be03}. Its closeness to the $D^0\bar D^{*0}$ threshold 
suggests that it may not be a standard charmonium, but rather a ``molecular" 
bound state of those mesons, of the type predicted by Tornqvist \cite{t91}.
(A review of experimental developments and theoretical questions can be
found in Ref.~\cite{br11}.) Clues to the nature of the $X(3872)$ have been 
sought in its decay modes, one of the most important of which is 
$X\rightarrow D^0\bar D^0\pi^0$ (see, for example, 
Refs.~\cite{bl07,kn09,hkn10}), but no definitive conclusion has yet been reached.
 
The fact that the $X(3872)$ lies within 1~MeV of a threshold makes it a 
suitable candidate for study using an effective field theory (EFT), similar to 
the ones that describe nucleon-nucleon scattering \cite{bvkrev,eprev}.
In particular, a nonrelativistic EFT including pion degrees of freedom, 
XEFT, was proposed by Fleming \textit{et al.}~\cite{fkmvk07}. This theory
is applicable to near-threshold states with unstable constituents.
A key ingredient is the hyperfine splitting between the $D^0$ and $D^{*0}$ which is 
$\Delta=M_{D^*}-M_D\simeq 142$~MeV. This means that the $D^*$ sits close 
to the $D\pi$ threshold, and hence has a very small width for strong decays. 
Fleming \textit{et al.}~therefore introduce the low-energy scale 
\begin{equation}\delta=\Delta-m_\pi\simeq 7\;\mbox{MeV}
\end{equation} 
and take $\delta/m_\pi$ as an expansion parameter for their EFT.

In this version of XEFT, $m_\pi$ and $\Delta$ are treated as high energy 
scales, corresponding to physics that has been integrated out. This provides 
no systematic justification for a further expansion in powers of the ratios 
of $m_\pi$ to the high-energy scales in the system, for example $m_\rho$ or 
the chiral scale $4\pi f_\pi$. In particular, as discussed at the end of 
Section II of Ref.~\cite{fkmvk07}, it does not justify an expansion in
$m_\pi/M_D$, which is a natural one to make and which simplifies the 
calculations. 

A further issue arises in diagrams with a $D\bar D\pi$ intermediate state 
when these are evaluated
at lowest order in this second expansion. At this order, the kinetic energies 
of the $D$ mesons are suppressed by $m_\pi/M_D$ compared to the kinetic energy 
of the pion. However neglecting these energies removes the constraint on the 
momenta of the $D$ mesons, giving a $D\bar D\pi$ threshold with a two-body 
structure, rather than the correct three-body form. To avoid this, Fleming 
\textit{et al.}~and subsequent authors \cite{bs10,jhj14} retain the $D$-meson 
kinetic energies when they evaluate the contributions from real pions. Although
doing so requires terms beyond leading order in their expansion, those authors
justified this on physical grounds, noting that the imaginary part of the 
self-energy is then consistent with the partial width for $X\rightarrow D\bar D\pi$
obtained by Voloshin from effective-range theory \cite{v04}.

In this note, we present a modified power counting for XEFT that avoids these 
problems. In contrast to Ref.~\cite{fkmvk07} we treat the pion mass and the 
hyperfine splitting $\Delta$  as of order $Q$, as in heavy-hadron chiral 
perturbation theory \cite{cas97}. We treat their difference $\delta$ as of 
order $Q^2$ in our counting. This reflects the near coincidence of the hyperfine
splitting of the $D$ mesons and the pion mass, which forms the basis for the 
original expansion of XEFT. In this new scheme, both the ratios $\delta/m_\pi$ 
and $m_\pi/M_D$ are of order $Q$, providing a common framework for both 
expansions. As we show below, it also ensures the correct behaviour at the 
$D\bar D\pi$ threshold.

Our starting point is the leading-order Lagrangian for the 
neutral mesons, as given in Eq.~(A.7) of Ref.~\cite{fkmvk07}:
\begin{equation}
\begin{split}
{\cal L}=&H^\dagger\left({\rm i}\partial_0+\frac{\nabla^2}{2M_D}\right)H
+{\bf H}^\dagger\cdot\left({\rm i}\partial_0+\frac{\nabla^2}{2M_D}-\Delta\right)
{\bf H}\cr
\noalign{\vspace{5pt}}
&+\,\frac{1}{2}\,\partial_\mu\pi_0\,\partial^\mu\pi_0-\frac{1}{2}\,m_\pi^2\,\pi_0^2
+\frac{g}{2 f_\pi}\left({\bf H}^\dagger\cdot\nabla\pi_0 H
+H^\dagger{\bf H}\cdot\nabla\pi_0\right),
\end{split}
\end{equation}
where the scalar field $H$ and the vector field ${\bf H}$ describe the $D^0$ 
and $D^{*0}$ mesons,
and the pion decay constant is $f_\pi=92.4$~MeV. This is supplemented by 
similar terms for the $\bar D^0$ and $\bar D^{*0}$ and a contact
interaction acting in the $C$-even combination of $D^0\bar D^{*0}$ 
and $\bar D^0D^{*0}$ channels. For generality, we treat the pion field as 
relativistic. 

The coupling strength of the pions to the charmed mesons is $g\simeq 0.6$ 
\cite{CLEO01}. The square of this is significantly smaller that for the 
corresponding coupling in the nuclear case (where $g_A=1.27$) and it suggests 
that treating pion exchange perturbatively may be a good approximation 
\cite{fkmvk07}. We therefore focus on the two diagrams shown in Fig.~1. 
The imaginary parts of these correspond to decay of the $X$ arising from 
coupling to a real or virtual $D\bar D^*$ (or $\bar DD^*$), followed by decay 
of the $\bar D^*$ (or $D^*$). The second diagram can be viewed as the contribution 
from interference between the $D\bar D^*$ and $\bar DD^*$ components of the $X$.

\begin{figure}[h]
\includegraphics[width=8cm,keepaspectratio,clip]{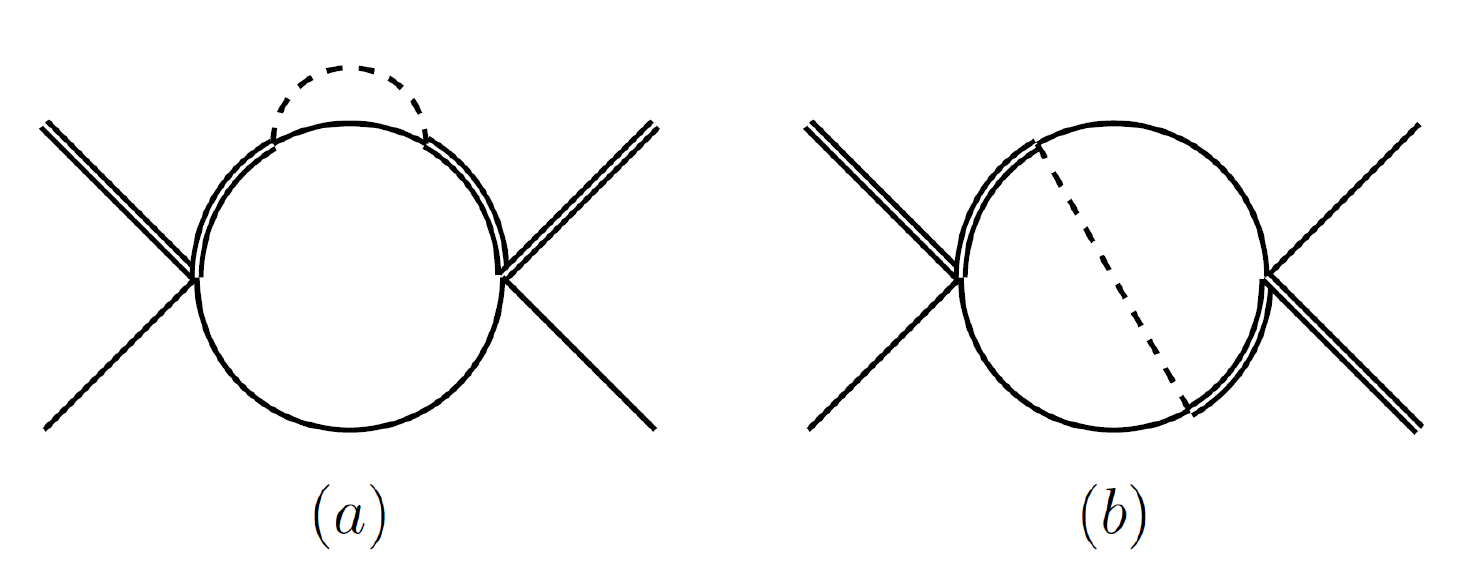}
\caption{The lowest-order pionic contributions to the self-energy of the $X$: 
(a) the self-energy of the $\bar D^*$, (b) pion exchange or, equivalently,
interference between $D\bar D^*$ and $\bar DD^*$ components of the wave function.
Solid lines represent $D$ mesons, double lines $D^*$ and dashed lines pions.}
\end{figure}

Using rotational invariance, the contribution
of diagram 1(a) can be written in the form
\begin{equation}
I^{(a)}_{ij}=\delta_{ij}\,\frac{1}{3}\,\left(\frac{g}{2\,f_\pi}\right)^2I^{(a)},
\label{eq:diag1a}
\end{equation}
where the loop integral is
\begin{equation}
 \begin{split}
I^{(a)}=&\,\int \frac{{\rm d}^4 k}{(2 \pi)^4} \int \frac{{\rm d}^4q}{(2 \pi)^4}\, 
\frac{q^2}{q_0^2-q^2-m_{\pi}^2+{\rm i}\epsilon}
\left(\frac{1}{k_0+\frac{E}{2}-\Delta-\frac{k^2}{2M_{D^*}}
+{\rm i}\epsilon}\right)^2\cr
\noalign{\vspace{5pt}}
&\qquad\qquad\qquad\qquad\times \,
\frac{1}{-k_0+\frac{E}{2}-\frac{k^2}{2M_D}+{\rm i}\epsilon}
\, \frac{1}{(k_0+q_0)+\frac{E}{2}-\frac{(k+q)^2}{2M_D}+{\rm i}\epsilon }\,.
\end{split}
\end{equation}
After integration over the energies $k^0$ and $q^0$, this becomes
\begin{equation}
\begin{split}
I^{(a)}=&\,\frac{1}{2}\,\int \frac{{\rm d}^3 k}{(2\pi)^3} 
\int \frac{{\rm d}^3 q}{(2 \pi)^3}\,
\frac{q^2}{\sqrt{q^2+m_\pi^2}}\left(\frac{1}{ E'-\frac{k^2}{2M_{r}}}\right)^2\cr
\noalign{\vspace{5pt}}
&\qquad\qquad\qquad\qquad\quad\times \,
\frac{1}{E'+\Delta-\frac{k^2}{2M_D}-\frac{(k+q)^2}{2M_D}-\sqrt{q^2+m_\pi^2}
+{\rm i}\,\epsilon}\,,
\label{eq:intdiag1a}
\end{split}
\end{equation}
where $M_r\simeq M_D/2$ is the reduced mass of the $D\bar D^*$ system 
and we have defined $E'=E-\Delta$, the total energy relative to the $D\bar D^*$ threshold.

We consider first the case where the $X$ lies below the $D\bar D^*$ threshold,
$E'<0$, and we expand the self-energy in powers of small scales. We treat the 
heavy particle momenta ($k$) as of order $Q$, as in nuclear EFTs \cite{bvkrev,eprev}. 
It follows that the energies ($k^2/2M_r$ and $E'$) are of order $Q^2$. 
The leading contribution to this integral comes from regions where the 
pion momentum $q$ is also of order $Q$. In this regime, the $D\bar D\pi$ energy 
denominator,
\begin{equation}
D(E')=E'+\Delta-\frac{k^2}{2M_D}-\frac{(k+q)^2}{2M_D}-\sqrt{q^2+m_\pi^2},
\label{eq:denomDDpi}
\end{equation}
is of order $Q$ and so it can be approximated by
\begin{equation}
D(E')\simeq\Delta-\sqrt{q^2+m_\pi^2}.
\end{equation}
The fact that the kinetic energies do not appear in the denominator at this order 
shows that 
this is the contribution of ``potential" pions in the terminology of Mehen and 
Stewart \cite{ms00}. The resulting contribution to the integral is of order $Q^2$, 
as expected for a leading two-loop diagram in a nonrelativistic EFT 
\cite{bvkrev,eprev}. 

The near cancellation between and $\Delta$ and $m_\pi$ leads to an enhanced
contribution to the integral from the region where $q^2\lesssim \Delta^2-m_\pi^2$.
In our proposed power counting, where $\Delta-m_\pi$ is taken to be of order $Q^2$, 
this implies that the pion momentum is 
\begin{equation}
q\lesssim\sqrt{\delta(\Delta+m_\pi)}\sim{\cal O}(Q^{3/2}).
\end{equation}
The $D\bar D\pi$ energy denominator is thus of order $Q^2$ and can be 
approximated here by 
\begin{equation}
D(E')\simeq E'+\delta-\frac{k^2}{2M_r}-\frac{q^2}{2m_\pi},
\end{equation}
showing that the pion is nonrelativistic in this regime. 
However, in contrast to the original expansion of XEFT \cite{fkmvk07}, 
both the pion and $D$-meson kinetic energies are of the same order in our 
counting. This ensures that the nonanalytic behaviour
at the $D\bar D\pi$ threshold has the correct three-body form.

At leading order in this counting, the imaginary part of the contribution of 
diagram 1(a) to the $X$ self-energy is proportional to
\begin{eqnarray}
{\rm Im}\bigl[I^{(a)}\bigr]&\simeq& -\,\frac{\pi}{2}\,\int \frac{{\rm d}^3 k}{(2 \pi)^3}
\left(\frac{1}{ E'-\frac{k^2}{2M_{r}}}\right)^2
\int \frac{{\rm d}^3 q}{(2 \pi)^3} \,\frac{q^2}{m_{\pi}}\,
\delta{\left(\frac{q^2}{2\,m_{\pi}}-E'-\delta+\frac{k^2}{2M_{r}}\right)}\cr
\noalign{\vspace{5pt}}
&=&-\,\frac{1}{8\pi^3}\left(\frac{m_\pi}{M_r}\right)^{3/2}
\int_0^{k_{\rm max}} k^2\,{\rm d}k\,\frac{[2M_r(E'+\delta)-k^2]^{3/2}}
{\bigl(E'-\frac{k^2}{2M_{r}}\bigr)^2}\,,
\label{eq:imX}
\end{eqnarray}
where
\begin{equation}
k_{\rm max}=\sqrt{2M_r(E'+\delta)}.
\end{equation}
The resulting imaginary part of the self energy is of order $Q^{7/2}$.

The region around the $D\bar D\pi$ threshold also makes a contribution of this 
same order, $Q^{7/2}$, to the real part of the self-energy. This may be of higher 
order than the contribution of the potential pions but it governs the 
nonanalytic behaviour at the threshold. This is relevant to studies of the 
quark-mass dependence of the $X(3872)$ \cite{ww13,befhmn13,jhj14} for analyses 
of lattice simulations of this state \cite{pl13}.

The situation is more complicated if the state lies above the  $D\bar D^*$ 
threshold and so $E'>0$. This also applies to calculations of the line shape 
for processes such as $B^+\rightarrow K^++X$ above this threshold \cite{bl07,hkn10}.
for these energies there is an additional enhancement to the integrals from the 
region around the $D\bar D^*$ threshold. In fact the expression in 
Eq.~(\ref{eq:imX}) for the
imaginary part diverges as a result of the double pole at $k^2=2M_r\,E'$. 
This is because, for energies close to the resonance, we cannot ignore the  
width of the $D^*$. Similar issues arise in the single-baryon system
at energies close to the pole  of the $\Delta$ resonance. As in the 
``$\delta$-counting" developed there \cite{pp03}, we include the
imaginary part of the self-energy to all orders in the $D^*$ propagator.

To leading order, the width of the $D^*$ is \cite{st98}
\begin{equation}
\Gamma_{D^*}=\frac{\sqrt{2}}{3\,\pi}\left(\frac{g}{2\,f_\pi}\right)^2 
[m_\pi(\Delta-m_\pi)]^{3/2},
\label{eq:wDstar}
\end{equation}
This is proportional to $(m_\pi\delta)^{3/2}$ and hence is of order $Q^{9/2}$ 
in our counting.
The width is thus much smaller than the typical values of the energy above the
$D\bar D^*$ threshold, which are of order $Q^2$. The $D^*$ propagator is enhanced
for energies within $\sim\Gamma_{D^*} $ of the pole. As discussed by Hanhart 
\textit{et al.}~\cite{hkn10}, the narrowness of this region means 
that we can neglect any energy dependence, and just replace the imaginary 
part of the self-energy by the on-shell width of the $D^*$.

This leads to the following expression for the imaginary part:
\begin{eqnarray}
 {\rm Im}\bigl[I^{(a)}\bigr]&\simeq& -\,\frac{\pi}{2}
\int \frac{{\rm d}^3 k}{(2 \pi)^3}\,
\frac{1}{\bigl(E'-\frac{k^2}{2M_{r}}\bigr)^2+\frac{1}{4}\Gamma_{D^*}^2}\,
\int \frac{{\rm d}^3 q}{(2 \pi)^3}\, \frac{q^2}{m_{\pi}}\,
\delta{\left(\frac{q^2}{2\,m_{\pi}}-E'-\delta+\frac{k^2}{2M_{r}}\right)}\cr
\noalign{\vspace{5pt}}
&=&-\,\frac{1}{8\pi^3}\left(\frac{m_\pi}{M_r}\right)^{3/2}
\int_0^{k_{\rm max}} k^2\,{\rm d}k\,\frac{[2M_r(E'+\delta)-k^2]^{3/2}}
{\bigl(E'-\frac{k^2}{2M_{r}}\bigr)^2+\frac{1}{4}\Gamma_{D^*}^2}\,.
\end{eqnarray}
The large contribution comes from a narrow region around $k^2=2M_rE'$
of width $\sim M_r\Gamma_{D^*}$ and it gives a result of order $Q$. 

This result can be simplified by noting that, in the region that gives the
dominant contribution, we can approximate the numerator of the integrand
using
\begin{equation}
2M_r(E'+\delta)-k^2\simeq 2M_r\delta+{\cal O}(Q^{9/2}).
\end{equation}
This leads to 
\begin{equation}
{\rm Im}\bigl[I^{(a)}\bigr]\simeq-\,\frac{\sqrt{2}}{4\pi^3}\,
\bigl(m_\pi\delta\bigr)^{3/2}
\int_0^{k_{\rm max}} k^2\,{\rm d}k\,\frac{1}
{\bigl(E'-\frac{k^2}{2M_{r}}\bigr)^2+\frac{1}{4}\Gamma_{D^*}^2}.
\end{equation}
The combination of low-energy scales multiplying the integral here is 
the same as appears in the width of the $D^*$, Eq.~(\ref{eq:wDstar}).
When we multiply by the coupling constants to get the full contribution 
from diagram 1(a), Eq.~(\ref{eq:diag1a}), we find that the 
result can expressed in the form
\begin{equation}
I^{(a)}_{ij}\simeq \delta_{ij}\,{\rm Im}\!\left[\frac{1}{2\pi^2}
\int k^2\,{\rm d}k\,\frac{1}{E'-\frac{k^2}{2M_{r}}+\frac{\rm i}{2}\,\Gamma_{D^*}}
\right]\,.
\end{equation}
This can be seen to be just the imaginary part of the self-energy in a theory 
without explicit pions but with an unstable $D^*$. Such a result should not be 
surprising, as the width of the $D^*$ is much smaller than that of the $X$, at 
least in the case that the $X$ lies above threshold and so can decay to the 
$D\bar D^*$ channel.  
The decay $D^*\rightarrow D\pi$ therefore occurs on a much longer timescale than 
that for $X\rightarrow D\bar D^*$. The width of the $X$ is thus independent of the 
details of the subsequent decay of the $D^*$. This explains the observations 
of Ref.~\cite{hkn10} that the result of the full calculation
for $X\rightarrow D\bar D\pi$ can be very well approximated by that for
$X\rightarrow D\bar D^*$, provided the energy is far enough above
the threshold, $E'\gg\Gamma_{D^*}$.

The pion-exchange or interference diagram, Fig.~1(b), can be treated in a very 
similar way. Integrating its contribution over $k_0$ and $q_0$ leads to two terms. 
One term has the energy denominator for the $D\bar D\pi$ intermediate state, 
Eq.~(\ref{eq:denomDDpi}). It therefore shows the same threshold enhancements as 
we have discussed above for the self-energy diagram. In particular, for an $X$ 
below the $D\bar D^*$ threshold, the threshold region contributes at order 
$Q^{7/2}$. In fact, at leading order in our counting, both diagrams lead to the 
same integral, Eq.~(\ref{eq:imX}), and so make equal contributions to the decay 
of the $X$. The second term arises from a virtual $D^*\bar D^* \pi$ state, which 
has a much higher threshold. For the energies considered here, this term just 
contributes to the ``potential" pion part of the self-energy.

For energies above the $D\bar D^*$ threshold, the additional enhancements just discussed come into play. The charge symmetry of the $X$ wavefunction might
suggest that both the $D^*$ self-energy and pion exchange should be included
to all orders, requiring a full three-body treatment of the $D\bar D\pi$ system 
as in Ref.~\cite{bhfkkn11}. However there is an important difference between 
the diagrams. The momentum $q$ transferred by the pion means that one of the 
$\bar DD^*$ energy denominators in Eq.~(\ref{eq:intdiag1a}) should be 
replaced by $E'-(k+q)^2/2M_{r}$. This separates the poles of the two 
$\bar DD^*$ denominators by ${\bf k}\cdot{\bf q}/M_r$, which is of order 
$Q^{5/2}$. For this diagram, there is no potential
double pole in the $k$ integral and so the enhancement in the $D^*$ pole 
region is not as strong. This is consistent with the numerical estimates
of the interference term by Hanhart \textit{et al.}~\cite{hkn10},
who found it made relatively small contributions above the $D\bar D^*$ 
threshold. All of this indicates that pion exchange can still be treated 
perturbatively in this region.

In summary: we have proposed a new power counting for XEFT, the effective field 
theory for a near-threshold state with unstable constituents. Like the counting 
originally proposed by Fleming \textit{et al.}~\cite{fkmvk07}, this leads to
an expansion in $\delta$, the difference between the hyperfine splitting of 
the $D$ mesons and the pion mass. However by counting this difference as order
$Q^2$, we are able to combine this in a single framework with an expansion 
in $m_\pi$, which we count in the usual way as of order $Q$. In this approach, 
both the pion and heavy-meson kinetic energies are 
of the same order and so amplitudes at the $D\bar D\pi$ threshold show the correct
three-body behaviour. This expansion allows us to recover within an EFT a number
of results that were previously obtained using physically-motivated approximations
\cite{v04,fkmvk07,hkn10}. It can provide a systematic framework for extending them
to higher orders.

\section*{Acknowledgments}

We are grateful to C. Hanhart, V. Lensky and J. McGovern for helpful
discussions. MA was supported by a scholarship
from the King Abdulaziz City for Science and Technology,
Saudi Arabia. MCB was supported in part by the UK STFC under grant ST/J000159/1 
and by the EU Integrated Infrastructure Initiative HadronPhysics3.

\end{document}